\documentclass[10pt, prd,twocolumn, nofootinbib,preprint,superscriptaddress]{revtex4}
\pdfoutput=1
\usepackage{amsmath,amssymb}
\usepackage{epsfig}
\usepackage{graphicx}
\usepackage[usenames,dvipsnames]{color}
\usepackage{subfigure}
\usepackage{slashed}
\usepackage[colorlinks,citecolor=blue]{hyperref}
\usepackage{color}

\begin{document}	
	\title{Scotogenic $U(1)_{L_{\mu}-L_{\tau}}$ origin of $(g-2)_\mu$, W-mass anomaly and 95 GeV excess}

	\author{Debasish Borah}
	\email{dborah@iitg.ac.in}
	\affiliation{Department of Physics, Indian Institute of Technology Guwahati, Assam 781039, India}
	
	\author{Satyabrata Mahapatra}
	\email{satyabrata@g.skku.edu}
	\affiliation{Department of Physics and Institute of Basic Science, Sungkyunkwan University, Suwon 16419, Korea}
	
	\author{Partha Kumar Paul}
	\email{ph22resch11012@iith.ac.in}
	\affiliation{Department of Physics, Indian Institute of Technology Hyderabad, Kandi, Sangareddy 502285, Telangana, India}
	
	\author{Narendra Sahu}
	\email{nsahu@phy.iith.ac.in}
	\affiliation{Department of Physics, Indian Institute of Technology Hyderabad, Kandi, Sangareddy 502285, Telangana, India}
	\begin{abstract}
		We study a scotogenic extension of the minimal gauged $L_{\mu}-L_{\tau}$ model including three right-handed singlet fermions and a 
		scalar doublet all odd under an in-built $Z_2$ symmetry to explain the anomalous magnetic moments of the muon, CDF-II W-mass anomaly, and the 95 GeV excess reported by the CMS collaboration. While the minimal model can successfully explain the muon $(g-2)$ and CDF-II W-mass anomalies, the required diphoton signal strength for the 95 GeV scalar, together with that of the SM Higgs, can not be obtained in the minimal model. The same can, however, be explained by incorporating one additional scalar doublet whose only role is to contribute radiatively to diphoton decay modes of the light, neutral scalars. Due to the scotogenic extension, the model remains consistent with the observed properties of light neutrinos and dark matter in the Universe. 
	\end{abstract}
	\maketitle
	\noindent
	\section{Introduction}
	The muon anomalous magnetic moment, $a_\mu$ = $(g - 2)_\mu/2$ has been measured recently by the E989
	experiment at Fermilab showing a discrepancy  with respect to the theoretical prediction of the Standard
	Model (SM) ($
	a^{\rm SM}_\mu = 116 591 810(43) \times 10^{-11}
	$) \cite{Abi:2021gix}. The 2021 analysis by the Muon g-2 collaboration, in combination with the previous Brookhaven results, led to a $4.2\sigma$ discrepancy. More recent analysis by the same collaboration \cite{Muong-2:2023cdq} has led to $\Delta a_\mu = a^{\rm exp}_\mu - a^{\rm SM}_\mu = 249 (48)\times 10^{-11}$, a discrepancy of $5.1\sigma$ CL. Here, it is worth noting that, due to the non-perturbative character of the low energy strong interaction, the uncertainty in $a^{\rm SM}_\mu$ is mostly dominated by hadronic vacuum polarization (HVP) contributions. These contributions are calculated from data-driven approaches, utilizing measured $e^{+} e^{-} \to {\rm hadrons}$ data or from Lattice QCD. Results from various lattice groups are combined using a conservative procedure to give an average value to a leading order (LO) as $a^{\rm LO~HVP}_\mu=711.6(18.4)\times10^{-10}$\cite{Aoyama:2020ynm}. The most recent result from Lattice QCD with higher precision is from BMW-20 analysis~\cite{Borsanyi:2020mff} which gives $a^{\rm LO~HVP}_\mu=707.5(5.5)\times10^{-10}$. Similarly, earlier measurements of HVP using $e^{+}e^{-} \to \pi^{+}\pi^{-}$ gives $a^{\rm HVP}_\mu= 6845(40)\times 10^{-11}$~\cite{Aoyama:2020ynm} and the same has been measured with greater precision by the recent CMD-3 experiment~\cite{CMD-3:2023alj}. While these observations contribute to alleviating the discrepancy between the experimental value of $a_\mu$ and its standard model (SM) prediction, they are not definitive, and the prospect of new physics beyond the standard model (BSM) being the cause of this anomaly still persists. Review of such theoretical explanations for muon $(g-2)$ can be found in \cite{Jegerlehner:2009ry, Lindner:2016bgg, Athron:2021iuf}.

	Similar anomalies have also been reported by collider experiments. The CMS experiment at the large hadron collider (LHC) has recently reported evidence for a neutral scalar of 95 GeV mass decaying into a pair of photons \cite{CMS:2018cyk, CMS:2022goy, CMS:2023yay} at $2.9\sigma$ CL which is also supported by the LEP data \cite{LEPWorkingGroupforHiggsbosonsearches:2003ing}. This excess is also supported by the ATLAS data but with a slightly lower statistical significance. This has led to several BSM explanations appeared in the literature \cite{Cao:2016uwt,Haisch:2017gql,Fox:2017uwr,Liu:2018xsw,Biekotter:2019kde,Cline:2019okt,Choi:2019yrv,Kundu:2019nqo,Cao:2019ofo,Biekotter:2020cjs,Abdelalim:2020xfk,Heinemeyer:2021msz,Biekotter:2022jyr,Iguro:2022dok,Biekotter:2023jld,Azevedo:2023zkg,Biekotter:2023oen,Escribano:2023hxj,Belyaev:2023xnv,Aguilar-Saavedra:2023vpd,Bhattacharya:2023lmu,Ashanujjaman:2023etj,Cao:2023gkc,Iguro:2022fel}. The other collider anomaly is the updated measurement of the W-boson mass $M_W = 80433.5 \pm 9.4$ MeV \cite{CDF:2022hxs} by CDF collaboration at Fermilab using the data corresponding to 8.8 ${\rm fb}^{-1}$ integrated luminosity collected at the CDF-II detector of Fermilab Tevatron collider. This updated value has $7\sigma$ discrepancy with the SM expectation ($M_{W}=80357\pm6$ MeV). This has already been explained in the context of different BSM scenarios in the literature. For instance see~\cite{Babu:2022pdn,Fan:2022dck,Asadi:2022xiy,Strumia:2022qkt,Arcadi:2022dmt,Borah:2022zim,Borah:2022obi,Mahapatra:2023dbr,Senjanovic:2022zwy,Cherchiglia:2023utd,Cheng:2022aau}.

	Motivated by these, we consider a popular BSM scenario based on the gauged $L_{\mu}-L_{\tau}$ symmetry, which is anomaly-free \cite{He:1990pn, He:1991qd, Foot:1990mn}.  In the minimal $U(1)_{L_\mu-L_\tau}$ model, three right-handed neutrinos (RHNs) are introduced in order to generate light neutrino masses at the tree level \cite{Patra:2016shz,Borah:2020jzi,Borah:2021jzu}. However, in this paper, we intend to explain neutrino mass in a scotogenic fashion \cite{Tao:1996vb, Ma:2006km} by imposing an additional $Z_2$ symmetry under which the RHNs and a scalar doublet, $\eta$ are odd. The unbroken $Z_2$ symmetry guarantees a stable dark matter (DM) candidate. A $L_{\mu}-L_{\tau}$ extension of the scotogenic model was discussed earlier in the context of DM and muon $(g-2)$ in \cite{Baek:2015fea, Borah:2021khc}. We will show that the minimal scotogenic $L_{\mu}-L_{\tau}$ model is consistent with DM, neutrino mass while explaining the muon $(g-2)$ and W-mass anomaly. However, the explanation of the CMS 95 GeV excess with the observed Higgs signal strength requires the addition of one more scalar doublet. While we do not discuss other phenomenology of this additional scalar doublet in the present work, it can be motivated from neutrino mass point of view in the absence of multiple generations of RHN \cite{Escribano:2020iqq}.
	
	This paper is organized as follows. In section \ref{sec:minimal model}, we briefly discuss the minimal scotogenic $L_{\mu}-L_{\tau}$ model followed by the explanation of CDF-II W-mass anomaly in section \ref{sec:wmass} and  CMS 95 GeV excess in section \ref{sec:95gev}. We discuss the details of neutrino mass generation in section \ref{sec:neutrinomass} and muon $(g-2)$ anomaly in section \ref{sec:muong-2}. In section \ref{sec:darkmatter}, we discuss the DM phenomenology and finally conclude in section \ref{sec:conclusion}. 
	
	\noindent
	\section{The Minimal Model}
	\label{sec:minimal model}

	\begin{table}[h!]
		\small
		\begin{center}
			\begin{tabular}{||@{\hspace{0cm}}c@{\hspace{0cm}}|@{\hspace{0cm}}c@{\hspace{0cm}}|@{\hspace{0cm}}c@{\hspace{0cm}}|@{\hspace{0cm}}c@{\hspace{0cm}}||}
				\hline
				\hline
				\begin{tabular}{c}
					{\bf ~~~~ Gauge~~~~}\\
					{\bf ~~~~Group~~~~}\\ 
					\hline
					
					$SU(2)_{L}$\\ 
					\hline
					$U(1)_{Y}$\\ 
					\hline
					$U(1)_{L_\mu-L_\tau}$\\ 
					\hline
					$Z_2$\\ 
				\end{tabular}
				&
				&
				\begin{tabular}{c|c|c}
					\multicolumn{3}{c}{\bf Fermion Fields}\\
					\hline
					~~~$N_e$~~~& ~~~$N_{\mu}$~~~ & ~~~$N_{\tau}$~~~ \\
					\hline
					$1$&$1$&$1$\\
					\hline
					$0$&$0$&$0$\\
					\hline
					$0$&$1$&$-1$\\
					\hline
					$-1$&$-1$&$-1$\\
				\end{tabular}
				&
				\begin{tabular}{c|c|c}
					\multicolumn{3}{c}{\bf Scalar Field}\\
					\hline
					~~~$\Phi_{1}$~~~& ~~~$\Phi_2$~~~ & ~~~$\eta$~~~\\
					\hline
					$1$&$1$&$2$\\
					\hline
					$0$&$0$&$\frac{1}{2}$\\
					\hline
					$1$&$2$&$0$\\
					\hline
					$+1$&$+1$&$-1$\\
				\end{tabular}\\
				\hline
				\hline
			\end{tabular}
			\caption{New fields and their
				gauge charges in minimal scotogenic $U(1)_{L_{\mu}-L_{\tau}}$ model.}
			\label{tab1}
		\end{center}    
	\end{table}
	
	In the minimal scotogenic $L_{\mu}-L_{\tau}$ setup, the SM particle content is extended with three right-handed neutrinos ($N_e$, $N_\mu$, $N_\tau$), two singlet scalars ($\Phi_1$, $\Phi_2$) and one doublet scalar ($\eta$). For the generation of neutrino mass at one-loop as well as for a stable DM candidate, an additional discrete symmetry, $Z_2$, is imposed under which the RHN and doublet scalar $\eta$ are odd while all other particles are even. The light active neutrinos acquire their tiny mass from the scotogenic loop with the RHN and $\eta$ taking part in the loop. The charge assignment of the BSM fields under the $SU(2)_L \otimes U(1)_Y \otimes U(1)_{L_\mu-L_\tau}\otimes Z_2$ symmetry is shown in Table \ref{tab1}.

	The relevant fermion Lagrangian can be written as
	\begin{align}
	\mathcal{L} & \supseteq
	\overline{N_{e}} i \gamma^\mu \partial_\mu N_{e}+ \overline{N_{\mu}} i \gamma^\mu \mathfrak{D}_\mu N_{\mu}	+\overline{N_{\tau}} i \gamma^\mu \mathfrak{D}_\mu N_{\tau} 
	\nonumber\\&
	- \frac{M_{ee}}{2} \overline{N^C_e} N_e - M_{\mu \tau} \overline{N^C_{\mu}} N_{\tau}-Y_{e\mu} \Phi^{\dagger}_1 \overline{N^C_e} N_\mu  \nonumber\\&- Y_{e\tau} \Phi_1 \overline{N^C_e} N_\tau -Y_{\mu} \Phi^{\dagger}_2 \overline{N^C_\mu} N_\mu -Y_{\tau} \Phi_2 \overline{N^C_\tau} N_\tau\nonumber \\&-Y_{De} \overline{L}_e \tilde{\eta} N_e-Y_{D\mu} \overline{L}_\mu \tilde{\eta} N_\mu-Y_{D \tau} \overline{L}_\tau \tilde{\eta} N_\tau \nonumber\\&- Y_{le} \overline{L}_e H e_R - Y_{l\mu} \overline{L}_\mu H \mu_R
	- Y_{l\tau} \overline{L}_\tau H \tau_R +{\rm h.c.}
	\label{eq:yuklag} \end{align}
	where $H$ is the SM Higgs doublet with vanishing $L_{\mu}-L_{\tau}$ charge, $\tilde{\eta}=i\tau_2 \eta^{*}$ and the covariant derivative $\mathfrak{D}_\mu$ is given as 
	\begin{equation}
	\begin{aligned}
	\mathfrak{D}_\alpha& \equiv \partial_\alpha - ig_{\mu \tau} Y_{\mu \tau} (Z_{\mu \tau})_\alpha\,.
	\end{aligned}
	\end{equation}
	with $Y_{\mu \tau}$ being the corresponding $U(1)_{L_\mu-L_\tau}$ charge. The new gauge kinetic terms in the Lagrangian are
	\begin{equation}
	\mathcal{L}_{\rm Gauge} = -\frac{1}{4} B_{\mu\nu}B^{\mu \nu}-\frac{1}{4} (Z_{\mu \tau})_{\mu\nu}(Z_{\mu \tau})^{\mu \nu} - \frac{\epsilon}{2} (Z_{\mu \tau})_{\mu\nu} B^{\mu \nu}
	\end{equation}
	where, $\epsilon$ is the kinetic mixing parameter between the $U(1)_{L_\mu - L_\tau}$ and $U(1)_Y$.

	The scalar sector 
	Lagrangian can  be written as:
	\begin{equation}
	\mathcal{L}_{scalar}= |\mathcal{D}_\mu{H}|^2+|\mathcal{D}_\mu{\eta}|^2+|\mathfrak{D}_\mu{\Phi}_{i}|^2 - V(H,\Phi_{i}, \eta)
	\end{equation}
	where $i=1,2$ and the covariant derivative $\mathcal{D}_\mu$ is given by
	\begin{equation}
	\begin{aligned}
	\mathcal{D}_\mu& \equiv \partial_\mu  - i\frac{g}{2}\tau.W_\mu - ig'\frac{Y}{2}B_\mu.
	\end{aligned}
	\end{equation}
	
	The scalar potential respecting the imposed symmetry can be written as
	
	\begin{align}
	V(H, \Phi_{i},\eta)&= -{\mu^2_H} \left(H^\dagger H \right) + \lambda_H \left(H^\dagger H \right)^2 -{\mu^2_{\Phi_i}} ({\Phi^\dagger_{i}} \Phi_{i} )\nonumber\\
	&+ \lambda_{\Phi_i} ({\Phi^\dagger_{i}} \Phi_{i} )^2 \nonumber  + \lambda_{H \Phi_i} (H^\dagger H)({\Phi^\dagger_{i}} \Phi_{i} )\,\nonumber\\&+m^2_{\eta}(\eta^\dagger \eta)+\lambda_2 (\eta^\dagger \eta)^2+\lambda_{3} (\eta^{\dagger} \eta) (H^{\dagger} H)\nonumber \\& + \lambda_4 (\eta^{\dagger} H) (H^{\dagger} \eta) 
	+ \dfrac{\lambda_{5}}{2} [ (H^{\dagger}\eta)^2+ (\eta^{\dagger}H)^2]\nonumber \\&
	+\lambda_{\eta \Phi_i} (\eta^\dagger \eta)(\Phi_{i}^\dagger \Phi_{i}) + \lambda_{\Phi_1\Phi_2}(\Phi_1^\dagger \Phi_1)(\Phi_2^\dagger \Phi_2) \nonumber\\& + [\mu \Phi^2_1  \Phi^\dagger_2+{\rm h.c.}]\label{eq:pot}
	\end{align}
	
	As the neutral component of the Higgs doublet $H$ breaks the electroweak gauge symmetry, the singlets $\Phi_1$ and $\Phi_2$, upon obtaining non-zero vacuum expectation values (VEV), break the $L_\mu - L_\tau$ gauge symmetry spontaneously. Here, it is worth mentioning that the non-trivial mixing of light neutrinos is a result of the structure of the RHN mass matrix, which is generated by the scalar singlets $\Phi_1$ and $\Phi_2$. Interestingly, one of these scalars responsible for breaking the $L_\mu-L_\tau$ symmetry and getting mixed with other $Z_2$-even neutral scalars is assumed to have a mass of 95 GeV and is responsible for the 95 GeV excess observed in collider experiments. We discuss these details in the subsequent sections. 
	
	The VEV alignments of the scalars are given as,
	\begin{eqnarray}
	H=\begin{pmatrix}
	0\\
	\frac{v+h}{\sqrt{2}}
	\end{pmatrix}, \Phi_1=\frac{1}{\sqrt{2}}(\phi_1+v_1),\Phi_2=\frac{1}{\sqrt{2}}(\phi_2+v_2).
	\end{eqnarray}		
	The mass squared matrix for the $Z_2$-even neutral scalars in the basis $(h,\phi_1,\phi_2)^T$ can be written as:
	{\small
		\begin{eqnarray}
		\left(
		\begin{array}{ccc}
		2 \lambda_H v^2 & \lambda_{H\Phi_1}v v_1  & \lambda_{H\Phi_2}v v_2 \\
		\lambda_{H\Phi_1}v v_1 & 2 \lambda_{\Phi_1} v_1^2 &v_1(2\lambda_{\Phi_1\Phi_2} v_2+\sqrt{2} \mu_{12}) \\
		\lambda_{H\Phi_2}v v_2 &v_1(2\lambda_{\Phi_1\Phi_2} v_2+\sqrt{2} \mu_{12}) & 2 \lambda_{ \Phi_2} v_2^2 -\frac{\mu_{12}v^2_1}{\sqrt{2}v_2} \\
		\end{array}
		\right).\nonumber\\
		\label{eq:scalarmassmat}
		\end{eqnarray}}	
	The new gauge boson, $Z_{\mu\tau}$, obtains mass after the $L_\mu-L_\tau$ symmetry is broken and is given as,
	\begin{eqnarray}
	M_{Z_{\mu\tau}}=g_{\mu\tau}\sqrt{v_1^2+4v_2^2}.
	\label{eq:mzmt}
	\end{eqnarray}
	This new gauge coupling $g_{\mu\tau}$ and gauge boson mass $M_{Z_{\mu\tau}}$ play crucial roles in explaining the muon $(g-2)$ anomaly. 	
	
	\section{W-Mass Anomaly}
	\label{sec:wmass}

	The mass of the $W$-boson is intricately computed within the robust framework of the SM, drawing upon precisely measured input parameters. These parameters, which encapsulate the fine-structure constant ($\alpha$), the Fermi constant ($G_F$), and the mass of the $Z$-boson ($M_Z$), play pivotal roles in the calculations. Their numerical values are derived from extensive experimental measurements and are given as
	\cite{ParticleDataGroup:2020ssz} 
	\begin{eqnarray}\label{precise}
	\alpha^{-1}_{\rm em} &=& 137.035999084(51) \,, \,\,\delta \alpha_{\rm em} / \alpha_{\rm em} = 0.4 \times 10^{-9};\nonumber \\
	G_F &=& 1.1663787 \times 10^{-5} \, {\rm GeV}^{-2}, \,\, \delta G_F/G_F = 0.4 \times 10^{-5}; \nonumber\\
	M_{Z}&=& 91.1876 \pm 0.0021 \, {\rm GeV}, \,\, \delta M_{Z}/M_{Z} = 2.5\times 10^{-5}.\nonumber\\
	\end{eqnarray}
	It is through these well-established and rigorously determined input parameters that the mass of the $W$-boson is precisely ascertained theoretically as~\cite{Hollik:1988ii, Nagashima:2010jma, Borah:2022zim}
	\begin{equation}
	M_W^2 \left(1-\frac{M_W^2}{ M_Z^2}\right) = \frac{\pi \alpha_{\rm em}}{\sqrt{2} G_F}(1+\Delta r)
	\end{equation}
	where $\Delta r$ represents the contributions from the quantum corrections, and from the above equation, $M_W$ can be calculated as
	\begin{equation}
	M_W^2 = \frac{M_Z^2}{2}\left[ 1 + \sqrt{1- \frac{4 \pi \alpha_{\rm em}}{ \sqrt{2} G_F M_Z^2}(1+\Delta r)} \right]~.
	\label{MW}
	\end{equation}

	This radiative contribution $\Delta r$ to the $W$-mass can be written as \cite{Hollik:1988ii}
	\begin{equation}
	\Delta r = \Delta \alpha_{\rm em} -\frac{\cos^2\theta_W}{\sin^2\theta_W} \Delta \rho + \Delta r_{1}
	\end{equation}
	with $\theta_W$ being the Weinberg angle. The primary factors influencing $\Delta r$ can be attributed to two main components. Firstly, there is the contribution from pure quantum electrodynamics (QED) correction, specifically the alteration of the fine structure constant $\Delta \alpha_{\rm em}$ as it evolves from $q^2= 0$ to $q^2 =M^2_Z$. The second component is $\Delta \rho$, which represents the vacuum polarization effect of the gauge boson through the top-bottom fermion loop.
	The change in the fine structure constant $\Delta \alpha_{\rm em}$ is
	\begin{eqnarray}
	\Delta \alpha_{\rm em} &=& \frac{\alpha_{\rm em}(M^2_Z)-\alpha_{\rm em}(0)}{\alpha_{\rm em}(0)}\nonumber\\
	&=& -\frac{\alpha_{\rm em}(0)}{3\pi} \sum_{m_f<M_Z} Q^2_f \Bigg[\frac{5}{3}-\ln(\frac{M^2_Z}{m^2_f})\Bigg]\nonumber\\
	&=& 0.05943 \pm 0.00011.
	\end{eqnarray}
	This stems from the renormalization of $\alpha$, primarily influenced by the contributions of light fermions.
	Likewise, $\Delta \rho$ represents the oblique correction arising from a significant contribution attributed to the top and Higgs loops, expressed as $({\cos^2\theta_W}/{\sin^2\theta_W}) \Delta \rho \simeq 0.03$. Beyond these, there are further contributions to $\Delta r$ originating from both vertex corrections and box diagrams, collectively denoted as $\Delta r_1= \Delta_{\rm Box+Vertex} \simeq 0.0064$.
	
	While utilizing the central values of the parameters provided in Eq.~\eqref{precise}, the calculated result is $\Delta r^{\rm}|_{\rm SM} = 0.0381387$. Consequently, employing Eq.~\eqref{MW}, this yields $M_W= 80.3564$ GeV. Notably, this value of $M_W$ is deviating by $7\sigma$ from the recently reported value by the CDF collaboration \cite{CDF:2022hxs}. To address this substantial discrepancy, quantum corrections influencing $\Delta r$ from its SM value may play a crucial role. We find that a value of $\Delta r = 0.033192$ aligns with the central value derived from the CDF measurement, i.e., $M_W = 80.4335$ GeV. Considering the potential modification of $\Delta r$ through oblique corrections, we identify the necessity for a new positive contribution to $\Delta \rho$, denoted as $\Delta \rho^{\rm req} = 0.00149014$, to account for the anomaly observed in the $W$-mass. This positive contribution can come from the self-energy correction of the W-boson with the new doublet scalar present in our setup. This additional contribution to self-energy correction $\Delta \rho$ and hence the $T$- parameter ($= {\Delta \rho}/{\alpha_{\rm em}}$) is given by \cite{Babu:2022pdn}: 
	\begin{eqnarray} 
	T  =\scriptstyle \dfrac{{\Theta}(m_{\eta^+}^2,m_{\eta_R}^2) + \Theta(m_{\eta^+}^2,m_{\eta_I}^2) 
		-\Theta(m_{\eta_R}^2,m_{\eta_I}^2)}{16\pi^2 \alpha_{\rm em}(M_Z) v^2 }\,, \label{eq:T}\nonumber\\
	\end{eqnarray}
	where the symmetric function $\Theta$ is given by
	\begin{equation} \label{Fdef}
	\Theta(x,y) \  \equiv \  \frac{1}{2}(x+y) -\frac{xy}{x-y}\ln\left(\frac{x}{y}\right)\,.
	\end{equation}
	
	In addition to the $T$-parameter contribution, the $S$-parameter can also modify the $W$-boson mass slightly. The $S$ parameter is given as,
	\begin{eqnarray}
	S=\frac{1}{12\pi}\log\bigg[\frac{M_{\eta_R}^2+M_{\eta_I}^2}{2M_{\eta^+}^2}\bigg].
	\end{eqnarray}
	The modified $W$-boson mass considering both these contributions is given by \cite{Grimus:2008nb, Borah:2022obi}
	\begin{align} 
	& M_W  \simeq M^{SM}_{W}\left[1- \dfrac{\alpha_{\rm em}(M_Z)(S-2~\cos^2{\theta_W}~ T)}{4(\cos^2{\theta_W}-\sin^2\theta_W)}\right]. \label{eq:MW-S}
	\end{align}	
	
	We observe that the alteration in the W-boson mass caused by the $S$ parameter is generally negligible, with the primary correction arising dominantly from the $T$ parameter.
	We investigate the parameter space within our model capable of elucidating the W-mass anomaly and the 95 GeV excess, as elaborated in the subsequent section, where the analysis ensures its consistency with all other phenomenological and experimental constraints.
	
	\section{95 GeV Excess}
	\label{sec:95gev}
	We now investigate whether our model can incorporate the experimental (CMS and ATLAS) indications of a BSM scalar at 95 GeV. In our setup, the $L_\mu-L_\tau$ symmetry is broken by the $Z_2$-even scalars $\Phi_1$ and $\Phi_2$. Once these scalars acquire VEV, they mix among themselves and with the SM Higgs, and we obtain three physical scalars as $H_1$, $H_2$, and $H_3$ in the particle spectrum. We identify $H_1$ as the SM Higgs boson with a mass of 125 GeV, $H_2$ as the 95 GeV scalar, and $H_3$ as a heavy scalar, which we assume to be heavy compared to the other two.
	The flavor eigenstates \{$h,\phi_1,\phi_2$\} and the mass eigenstates \{$H_1, H_2, H_3$\} of these scalars are related by an orthogonal transformation that diagonalizes the mass-matrix given in Eq.~\ref{eq:scalarmassmat}. This can be written as:
	\begin{widetext}
		\begin{eqnarray}
		\begin{pmatrix} h \\ \phi_1 \\ \phi_2 \end{pmatrix} &=&
		\left(
		\begin{array}{ccc}
		c_{12} c_{13} & c_{13} s_{12} & s_{13} \\
		-c_{12} s_{13} s_{23}-c_{23} s_{12} & c_{12} c_{23}-s_{12} s_{13} s_{23} & c_{13} s_{23} \\
		s_{12} s_{23}-c_{12} c_{23} s_{13} & -c_{12} s_{23}-c_{23} s_{12} s_{13} & c_{13} c_{23} \\
		\end{array}
		\right)
		\begin{pmatrix} H_1 \\ H_2 \\ H_3 \end{pmatrix} .\label{Eq:scalarmass}
		\end{eqnarray}
	\end{widetext}
	where we abbreviated $\cos\theta_{ij} = c_{ij}$ and $\sin \theta_{ij} = s_{ij}$. 
	For simplicity, we denote the scalars as follows,
	\begin{equation}
	H_1 \equiv H_{\rm SM},~  H_2 \equiv H_{95} ,~ H_3\equiv H_{\rm heavy}. 
	\end{equation}
	The signal strength for a particular process, denoted as $\mu$, is determined by the ratio of the observed number of events to the expected number of events. For a hypothetical SM like Higgs boson, denoted as $H_{95}$, with a mass of 95 GeV, the signal strength for the observed diphoton excess in our setup is expressed as,
	\begin{align}
	\mu^{\gamma\gamma}_{H_{95}} &=& \frac{\sigma(gg\to H_{95})}{\sigma_{\rm SM}(gg\to H_{95})} \times \frac{{\rm BR}(H_{95}\to\gamma\gamma)}{{\rm BR}_{\rm SM}(H_{95}\to\gamma\gamma)} \nonumber\\&=& \sin^2\theta_{12}\cos^2\theta_{13}\times \frac{{\rm BR}(H_{95}\to\gamma\gamma)}{{\rm BR}_{\rm SM}(H_{95}\to\gamma\gamma)} \, ,\label{eq:signalstre}
	\end{align}
	where ${\rm BR}(H_{95}\to\gamma\gamma)$ is the branching for 95 GeV scalar to two photons in our model, and ${\rm BR}_{\rm SM}(H_{95}\to\gamma\gamma)$ is the branching to two-photon state for a SM like Higgs state with a mass of 95 GeV. The existence of the inert doublet $\eta$ introduces extra loop contributions to the $\gamma\gamma$ decay channel of both $H_1$ and $H_2$, in addition to the contributions from the fermion loop and the $W$-boson loop. From Eqs \ref{eq:pot}, \ref{Eq:scalarmass}, the couplings of the $Z_2$-even scalars with the charged component of the inert doublet are given by
	\begin{eqnarray}
	C_{H_1 \eta \eta}&= c_{12} c_{13} \lambda_3 v + (-c_{23} s_{12} - 
	c_{12} s_{13} s_{23}) \lambda_{\eta \Phi_1} v_1  \nonumber\\&+ (-c_{12} c_{23} s_{13} + 
	s_{12} s_{23}) \lambda_{\eta \Phi_2} v_2 \, , \nonumber\\
	C_{H_2 \eta \eta}&= s_{12} c_{13} \lambda_3 v + (c_{23} c_{12} - 
	s_{12} s_{13} s_{23}) \lambda_{\eta \Phi_1} v_1  \nonumber\\&+ (-s_{12} c_{23} s_{13} + 
	c_{12} s_{23}) \lambda_{\eta \Phi_2} v_2 
	\, , \nonumber\\
	C_{H_3 \eta \eta}&= s_{13} \lambda_3 v + c_{13} s_{23}  \lambda_{\eta \Phi_1} v_1 + c_{13} c_{23} \lambda_{\eta \Phi_2} v_2.\label{eq:couplingss}
	\end{eqnarray}
	Thus the diphoton decay width of these $Z_2$-even scalars is given by~\cite{Djouadi:2005gi}
	\begin{widetext}
		\begin{equation}
		\Gamma(H_i\to \gamma\gamma) = \frac{G_F\alpha^2 m_{H_i}^3}{128\sqrt{2}\pi^3}\left| \sum_f N_c Q_f^2 g_{H_iff} \, A_{1/2}(\tau_f) + g_{H_iWW} \, A_1(\tau_W) +\frac{v}{2M^2_{\eta^+}}C_{H_i\eta\eta} \, A_0(\tau_\eta)\right|^2 \, 
		\label{diphotondec} 
		\end{equation}
	\end{widetext}
	
	where $\tau_k=m^2_{H_i}/4m^2_k$ and the loop functions $A_r(\tau)$ are defined in Appendix~\ref{app:loop}. $N_c$ is the color factor, $Q_f$ is the electric charge of the fermion, and $g_{H_i ff}$, $g_{H_i WW}$ are the corresponding couplings of the scalars with fermions and W-boson respectively.

	The CMS collaboration has reported the signal strength for this channel as $\mu^{\gamma\gamma}_{H_{95}}=0.33^{+0.19}_{-0.12}$. In the current analysis of the 95 GeV signal strength, crucial parameters include $\sin{\theta_{12}}$, $M_{\eta^+}$, $\lambda_{\eta \Phi_1}$, $\lambda_{3}$, and the right-handed neutrino (RHN) mixing angle ($\sin{\beta_{12}}$). As the production cross-section of $H_{95}$ is directly proportional to $\sin{\theta_{12} }$, if $\sin{\theta_{12}}$ is very small, the production cross-section becomes minimal, leading to insufficient signal strength. Conversely, with very large values of $\sin{\theta_{12}}$, $H_{95}$ predominantly decays into SM-charged fermions through mixing with SM Higgs, causing a reduction in the diphoton branching ratio. The mixing angles $\sin\theta_{13}$, and $\sin\theta_{23}$ can influence the signal strengths if they are sufficiently larger, as can be seen from Eq. (\ref{eq:couplingss}). Thus, achieving the desired 95 GeV signal strength necessitates precise tuning of the mixing angles. In addition, the mass of the charged component of the inert doublet $\eta^+$ cannot be arbitrarily large. As indicated in Eq. \eqref{diphotondec}, as $M_{\eta^+}$ increases, the decay rate of the 95 GeV scalar to the diphoton state decreases. In this context, it is essential to note that the necessary correction to the W-mass primarily stems from the $T$ parameter, which is highly sensitive to the mass difference between the neutral and charged components of the doublet scalar, which can be easily read from Eq. (\ref{eq:T}). Therefore, in seeking a shared parameter space that effectively accounts for both the 95 GeV signal strength and the W-mass anomaly, the mass of the doublet scalar becomes significantly constrained.
	
	For the numerical scan, we consider the couplings $\lambda_{3}$, $\lambda_{5}$, $\lambda_{\eta \Phi_1}$, $\lambda_{\eta \Phi_2}$, the charged scalar mass $M_{\eta^+}$, and the CP-even scalar mass $M_{\eta_{R}}$ as free parameters and determine $M_{\eta_{I}}$ and $\lambda_{4}$ using Eq. \eqref{eq:metar}, Eq. \eqref{eq:metai}, and Eq. \eqref{eq:metap}. These parameters are randomly varied within the following ranges: $M_{\eta_R}\in[45, 150]$ GeV, $\lambda_5\in [10^{-8},10^{-2}]$, $\lambda_3\in[0.01, \sqrt{4\pi}]$, $M_{\eta^+}-M_{\eta_R}\in[1, 100]$ GeV, $\lambda_{\eta\Phi_1}\in [0.01, 4\pi]$, $\lambda_{\eta\Phi_2}\in [0.01, 4\pi]$. The scalar mixing angles are varied within the range: $\sin\theta_{12}\in [10^{-4}, 0.7]$, $\sin\theta_{23}\in [10^{-4}, 0.7]$, and $\sin\theta_{13}\in [10^{-4}, 0.7]$.
	Concerning the RHN mixing angles, two of them, $\beta_{13}=0$ and $\beta_{23}=\pi/4$, are uniquely determined due to the symmetry. The only mixing angle that is randomly varied is $\sin\beta_{12}\in[10^{-4}, 0.7]$. Further details on this can be found in Appendix~\ref{app:rhnmixing}. Additionally, since the $H_2 \eta^+ \eta^-$ vertex depends on the VEV of $\Phi_1$, we randomly vary it as $v_1\in [20, 100]$ GeV and determine $v_2$ using Eq.~\eqref{eq:mzmt}. Throughout the whole analysis, we have ensured that all the couplings adhere to the perturbative bounds. We also impose the LEP limits on the doublet scalar as $M_{\eta_R}+M_{\eta_I} > M_{Z}$ and a conservative limit on the charged scalar $M_{\eta^+}>100$ GeV.
	
	\begin{figure}[h!]
		\centering
		\includegraphics[scale=0.45]{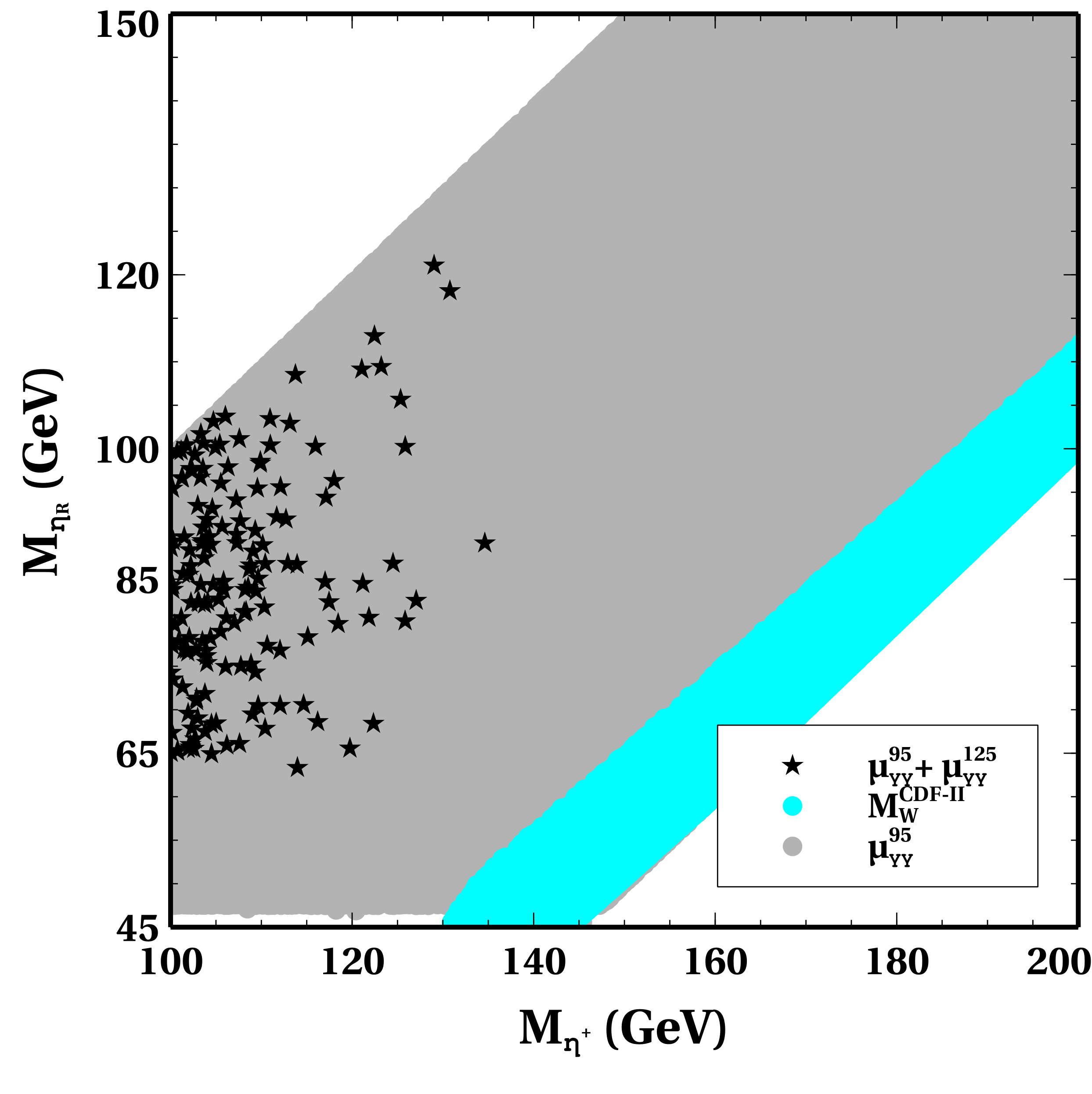}
		\caption{Parameter space satisfying CDF-II $W$-mass anomaly, 95 GeV excess, and SM Higgs diphoton signal strength in the plane of $M_{\eta_{R}}-M_{\eta^+}$ with one generation of $\eta$.}
		\label{fig:1genmetarvsmetap}
	\end{figure}
	In Fig. \ref{fig:1genmetarvsmetap}, we illustrate the parameter space that corresponds to the observed 95 GeV diphoton signal strength in the $M_{\eta_{R}}$ and $M_{\eta^+}$ plane, represented in grey. In the same plane, we depict the parameter space satisfying the W-mass anomaly in cyan. It is certain from Eq. (\ref{eq:T}) that in order to satisfy the required $W$ mass anomaly, the $T$ parameter demands larger mass splitting between the charged and the neutral components of the inert doublet. In this scenario, the mass difference between the charged and neutral components of the inert doublet falls within the range of $[80, 100]$ GeV, ensuring the required $T$ and $S$ parameters capable of explaining the CDF-II W-mass anomaly. However, it is important to note that for the validation of this parameter space, its consistency with the measured SM Higgs signal strength in the diphoton channel should be verified. The measured value of this signal strength is $\mu^{\gamma \gamma}_{H_{\rm SM}}= 1.1\pm 0.07$~\cite{ParticleDataGroup:2020ssz} in the $1\sigma$ limit. This is essential because the same newly introduced BSM particles also influence the SM Higgs branching ratio to the diphoton channel. 
	As depicted in Fig. \ref{fig:1genmetarvsmetap}, it is possible to simultaneously account for the 95 GeV signal strength and SM Higgs signal strength, which is shown by the black star-shaped points. This constrains the charged scalar mass approximately in the range of $\sim[100,135]$ GeV.
	However, it is important to note that no common parameter space satisfies all three criteria simultaneously, including the W-mass anomaly. As can be seen from Eq. (\ref{diphotondec}), the diphoton decay width decreases with an increase in the mass of $M_{\eta^+}$. This decrease in the decay width can be compensated in the signal strength by increasing the production cross-section, \textit{i.e.} increasing the mixing angle $\theta_{12}$ and decreasing $\theta_{13}$ in Eq. (\ref{eq:signalstre}). 	
		However, this, in turn, reduces the SM Higgs to diphoton signal strength as the production cross-section for the SM Higgs is proportional to $\cos\theta_{12} \cos\theta_{13}$.
	In models such as \cite{Escribano:2023hxj}, which involve one generation of $\eta$, the $Z_2$-even scalars decay into $\gamma \gamma$ via $\eta^+$ in the loop, in addition to the standard model channels. In contrast, in our gauged $U(1)_{L_\mu-L_\tau}$ scotogenic scenario, we have two additional decay modes for the scalars, namely, $H_i\rightarrow N_1 N_1$ and $H_i\rightarrow Z_{\mu\tau}Z_{\mu\tau}$. Even if $H_i\rightarrow N_1 N_1$ is tuned by suitable choices of Yukawa couplings, the latter can not be tuned arbitrarily due to tight constraints on $g_{\mu \tau}, M_{Z_{\mu \tau}}$ from $(g-2)_\mu$ as we discuss later. However, this mode can be tuned by choosing the $\theta_{13}$ and $\theta_{23}$ appropriately. The presence of these decay modes significantly reduces the scalar branching for the diphoton channel, resulting in a signal strength much less than the required. As we have seen in \ref{fig:1genmetarvsmetap}, the mass splitting between the charged and neutral components of $\eta$ is constrained to be in the range 80 to 100 GeV, and the SM Higgs signal strength is allowing the maximum $M_\eta^+$ to be 135 GeV, it is not possible to explain these two simultaneously. Clearly, if we can further decrease this mass splitting between the charged and neutral components of $\eta$, we can get a common parameter space satisfying all three criteria.
	
	\begin{figure}[h!]
		\centering
		\includegraphics[scale=0.45]{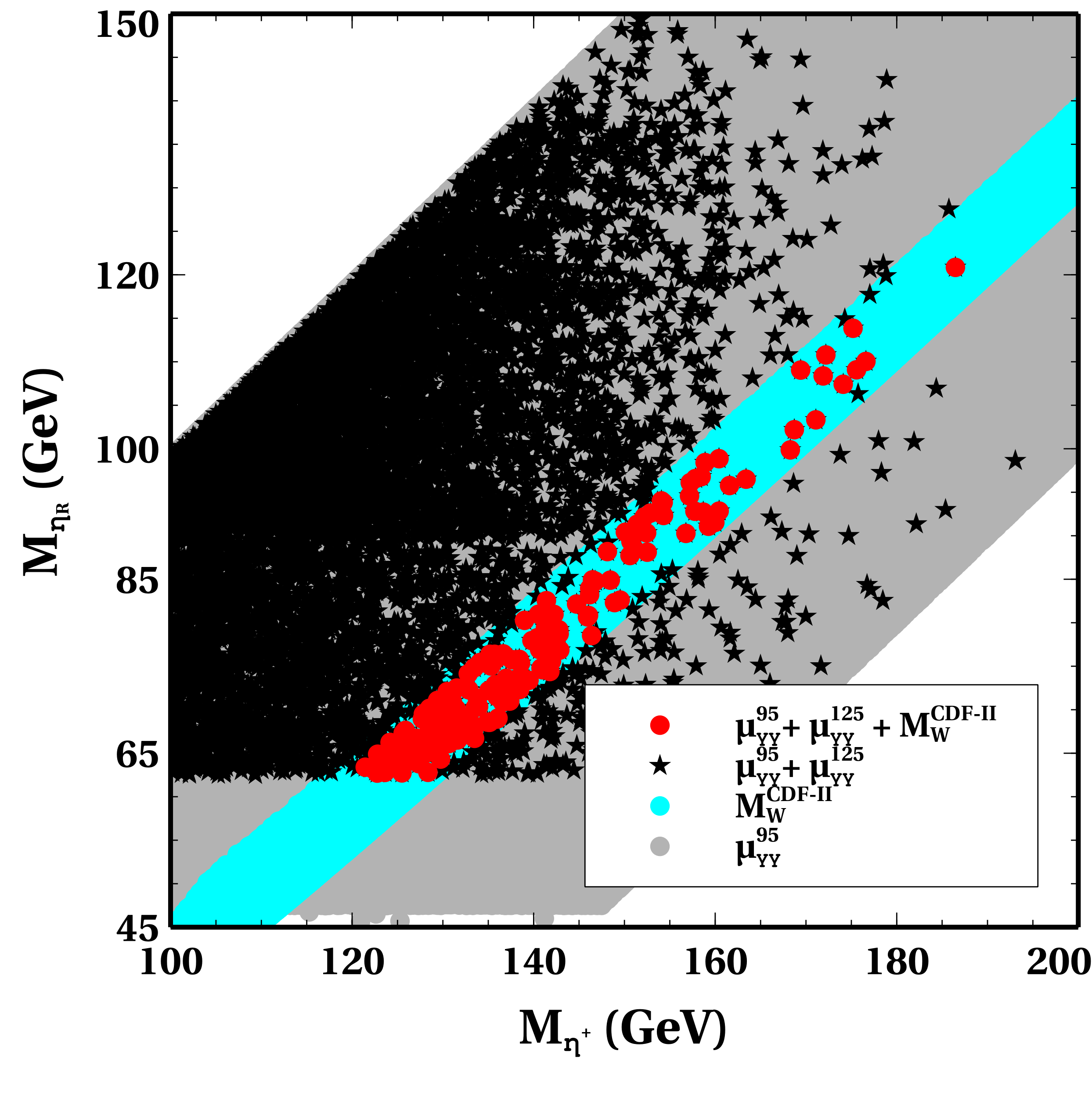}
		\caption{Parameter space satisfying CDF-II $W$-mass anomaly, 95 GeV excess, and SM Higgs signal strength in the plane of $M_{\eta_{R}}-M_{\eta^+}$ with two generations of $\eta$.}
		\label{fig:2genmetarvsmetap}
	\end{figure}
	To salvage the scenario, we introduce additional generation of $\eta$ to enhance the contribution to $H_{\text{SM}}\rightarrow\gamma \gamma$ for larger $M_{\eta^+}$ masses. This augmentation allows the explanation of the SM Higgs signal strength, the observed 95 GeV diphoton signal strength, and the W-mass anomaly simultaneously. The common parameter space satisfying the 95 GeV excess, SM Higgs signal strength, and W-mass anomaly simultaneously is presented in Fig. \ref{fig:2genmetarvsmetap}, depicted by the red colored points. The grey-colored points satisfy only the 95 GeV excess, while the black-colored stars satisfy the 95 GeV excess and SM Higgs signal strength, and the cyan-colored points satisfy the $W$-mass anomaly. As depicted in Fig. \ref{fig:2genmetarvsmetap}, the charged scalar mass is constrained to lie approximately in the range of $\sim[120,190]$ GeV. It is worth highlighting that with two generations of the inert doublet, the mass splitting between the charged and neutral components of the inert doublet is reduced to $[50, 70]$ GeV to explain the $W$ mass anomaly. Compared to Fig. \ref{fig:1genmetarvsmetap}, in Fig. \ref{fig:2genmetarvsmetap}, the W-mass favored region shifts upwards due to the decreased mass splitting. With two generations of $\eta$, the allowed mass range for $\eta^+$ based on the 95 GeV and SM Higgs to diphoton signal strength is $[120,190]$ GeV, while $M_{\eta_R}$ is in the range $[62.5, 120]$ GeV. For simplicity, we assume that the two doublets scalars are decoupled, meaning inter-generation mixings are absent.
	\begin{figure}[h!]
		\centering
		\includegraphics[scale=0.45]{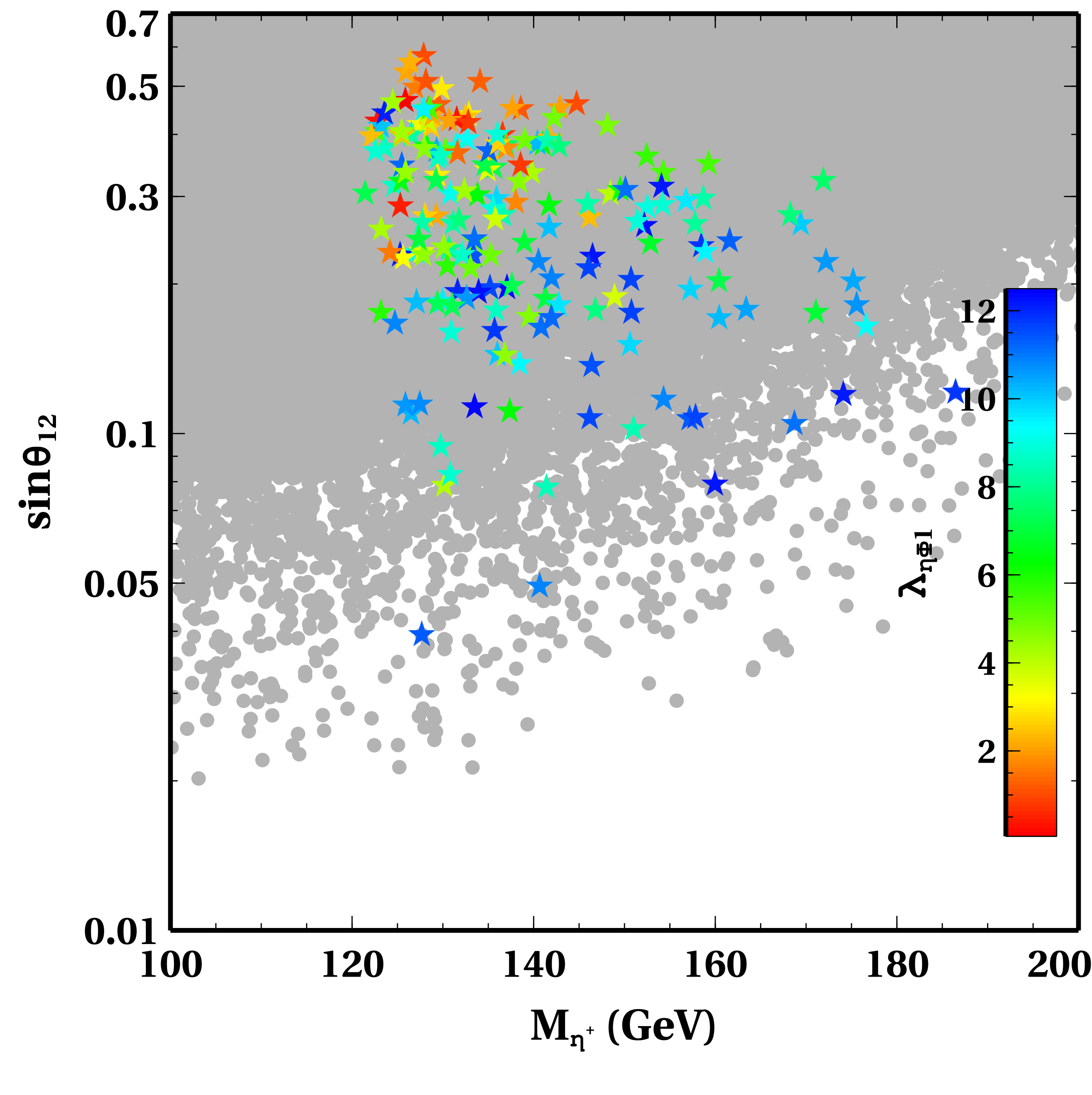}
		\caption{Allowed parameter space in the plane of $\sin\theta_{12}-M_{\eta^+}$ with two generations of $\eta$.}
		\label{fig:2gens12vsmetap}
	\end{figure}
	
	We also illustrate the parameter space in the $\sin\theta_{12}-M_{\eta^+}$ plane in Fig. \ref{fig:2gens12vsmetap}. The grey points depict the parameter space that satisfies only the 95 GeV diphoton signal strength and the SM Higgs diphoton signal strength. On the other hand, the colored points satisfy the 95 GeV scalar signal strength, the SM Higgs signal strength, and the $W$-mass anomaly. In this context, the scalar mixing angle $\sin\theta_{12}$ is confined within the range $[0.04, 0.6]$. From Eqs. \eqref{eq:signalstre}, \eqref{eq:couplingss}, \eqref{diphotondec}, we also see that as $\sin\theta_{12}$ decreases, the signal strength decreases. This can be compensated by increasing $\lambda_{\eta\Phi_1}$, as it solely enhances the diphoton decay width. This feature can be easily read from Fig. \ref{fig:2gens12vsmetap}. The scalar mixing angles, $\sin\theta_{13}$, and $\sin\theta_{23}$ are constrained in the ranges $[2\times10^{-4}, 4\times10^{-2}]$, and $[0.06, 0.3]$ respectively. The RHN mixing angle $\sin\beta_{12}$ can not be too large to explain both the 95 GeV excess and SM Higgs signal strength. This is because larger values of $\sin{\beta_{12}}$ lead to prominence of the $H_i\rightarrow N_1 N_1$ decay mode, resulting in a decrease in the corresponding diphoton branching.
	
	\section{Neutrino Mass}	
	\label{sec:neutrinomass}
	As previously discussed, the Majorana neutrino mass is induced at the one-loop level, involving the participation of particles from the dark sector in a manner akin to the scotogenic model~\cite{Tao:1996vb, Ma:2006km}. Due to the imposed symmetry that governs the interactions, it becomes evident from Eq.~\eqref{eq:yuklag} that both the charged lepton mass matrix and the Dirac Yukawa matrix of neutrinos exhibit a diagonal structure. Consequently, the non-trivial neutrino mixing can be generated by the structure of the RHN mass matrix, which is generated by the scalar singlet fields. The charged lepton mass matrix, neutrino Dirac Yukawa matrix, and the right-handed neutrino mass matrix, respectively, are given by
	\begin{align}
	M_\ell= \frac{1}{\sqrt{2}}\begin{pmatrix}
	Y_{e} v & 0 & 0\\
	0 & Y_{\mu}v & 0 \\
	0 & 0 & Y_{\tau} v
	\end{pmatrix},\,
	Y_D =\begin{pmatrix}
	Y_{De}       &  0    & 0  \\
	0     &  Y_{D\mu}    & 0  \\
	0     &  0    & Y_{D \tau}  
	\end{pmatrix},\,     
	\nonumber
	\end{align}
	\begin{equation}
	\label{mass_mat}
	M_R =\begin{pmatrix}
	M_{ee}      &  Y_{e\mu} v_1
	& Y_{e\tau} v_ 1 \\
	Y_{e\mu} v_ 1      &  Y_{\mu}
	v_2     & M_{\mu \tau}  \\
	Y_{e\tau} v_1     &  M_{\mu \tau}    &
	Y_{\tau} v_2  
	\end{pmatrix}\,.
	\end{equation}
	The neutrino mass arising at one-loop is given by: \cite{Ma:2006km, Merle:2015ica}
	\begin{align}
	(M_{\nu})_{\alpha \beta} \ & = \ \sum_k \frac{Y_{\alpha k}Y_{ k \beta} M_{k}}{32 \pi^2} \left[L_k(M^2_{\eta_R})-L_k(M^2_{\eta_I})\right] \, ,
	\label{numass1}
	\end{align}
	where  $M_{\eta_R}^2=M_{\eta_I}^2+\lambda_5 v^2$ from Eq. \eqref{eq:metar} and \eqref{eq:metai},
	$M_k$ represents the mass eigenvalue associated with the right-handed neutrino mass eigenstate $N_k$ running in the internal line, with the indices $\alpha$ and $\beta$ ranging from 1 to 3, covering the three neutrino generations. The Yukawa couplings $Y_{\alpha k}$ in the given neutrino mass formula are obtained from the Dirac Yukawa couplings in Lagrangian \eqref{eq:yuklag} by transitioning to the diagonal basis of right-handed neutrinos after spontaneous symmetry breaking. The loop function $L_k(m^2)$ in neutrino mass formula \eqref{numass1} is defined as 
	\begin{align}
	L_k(m^2) \ = \ \frac{m^2}{m^2-M^2_k} \: \text{ln} \frac{m^2}{M^2_k} \, .
	\label{eq:Lk}
	\end{align}
	It is noteworthy that the mass difference between the neutral scalar and pseudo-scalar components of $\eta$ ($M^2_{\eta_R}-M^2_{\eta_I}=\lambda_5 v^2$) plays a pivotal role in generating a non-zero neutrino mass. From Eq. (\ref{numass1}), it is evident that, in the limit $\lambda_5\rightarrow0$, $M_{\eta_R}=M_{\eta_I}$, which implies that  $M_\nu=0$. For the analysis, the initial step involves diagonalizing $M_R$ and considering the physical basis of right-handed neutrinos $(N_1, N_2, N_3)$ with appropriate interactions. Employing the Casas-Ibarra (CI) parametrization \cite{Casas:2001sr} extended to the radiative seesaw model \cite{Toma:2013zsa}, we express the Yukawa coupling matrix that satisfies the neutrino oscillation data as follows:
	
	\begin{align}
	Y_{\alpha k} \ = \ \left(U D_\nu^{1/2} R^{\dagger} \Lambda^{1/2} \right)_{\alpha k} \, ,
	\label{eq:Yuk}
	\end{align}
	
	In this expression, $U$ is the PMNS matrix, and $R$ is an arbitrary complex orthogonal matrix satisfying $RR^{T}=\mathrm{I}$. The diagonal light neutrino mass matrix is denoted by $D_\nu = \textrm{diag}(m_1, m_2, m_3)$, and the diagonal matrix $\Lambda$ has elements given by:
	
	\begin{align}
	\Lambda_k \ & = \ \frac{2\pi^2}{\lambda_5}\zeta_k\frac{2M_k}{v^2} \, , \\
	\textrm {and}\quad \zeta_k & \ = \  \left(\frac{M_{k}^2}{8(M_{\eta_R}^2-M_{\eta_I}^2)}\left[L_k(M_{\eta_R}^2)-L_k(M_{\eta_I}^2) \right]\right)^{-1} \, . \label{eq:zeta}
	\end{align}
	
	\section{Muon (\lowercase{g}-2) Anomaly} 
	\label{sec:muong-2}
	
	The magnetic moment of the muon is expressed as:
	\begin{equation}\label{mug2}
	\overrightarrow{\mu_\mu}= g_\mu \left (\frac{q}{2m_\mu} \right)
	\overrightarrow{S}\,,
	\end{equation}
	where $g_\mu$ denotes the gyro-magnetic ratio, assuming a value of 2 for an elementary spin-$\frac{1}{2}$ particle with $m_\mu$ and $q$ denoting the mass and electro-magnetic charge of the muon, respectively. From the Dirac equation, the value of the $g_\mu$ at the tree level is found to be 2. However, within the framework of quantum field theory, minuscule corrections to this value emerge owing to the particle's interactions with virtual particles and quantum effects. These corrections are parameterized as $a_\mu=( g_\mu - 2)/2$.
	In our framework, the primary augmentation to the muon magnetic moment predominantly comes from the one-loop diagram facilitated by the $L_{\mu}-L_{\tau}$ gauge boson, denoted as $Z_{\mu \tau}$. The associated one-loop contribution is expressed as \cite{Brodsky:1967sr, Baek:2008nz}:
	\begin{equation}
	\Delta a_{\mu} = \frac{\alpha_{\mu\tau}}{2\pi} \int^1_0 dx \frac{2m^2_{\mu} x^2 (1-x)}{x^2 m^2_{\mu}+(1-x)M^2_{Z_{\mu \tau}}} 
	\end{equation}
	Here, $\alpha_{\mu\tau}$ is defined as $\alpha_{\mu\tau}=g^2_{\mu \tau}/(4\pi)$.
	
	\begin{figure}[h!]
		\centering
		\includegraphics[scale=0.45]{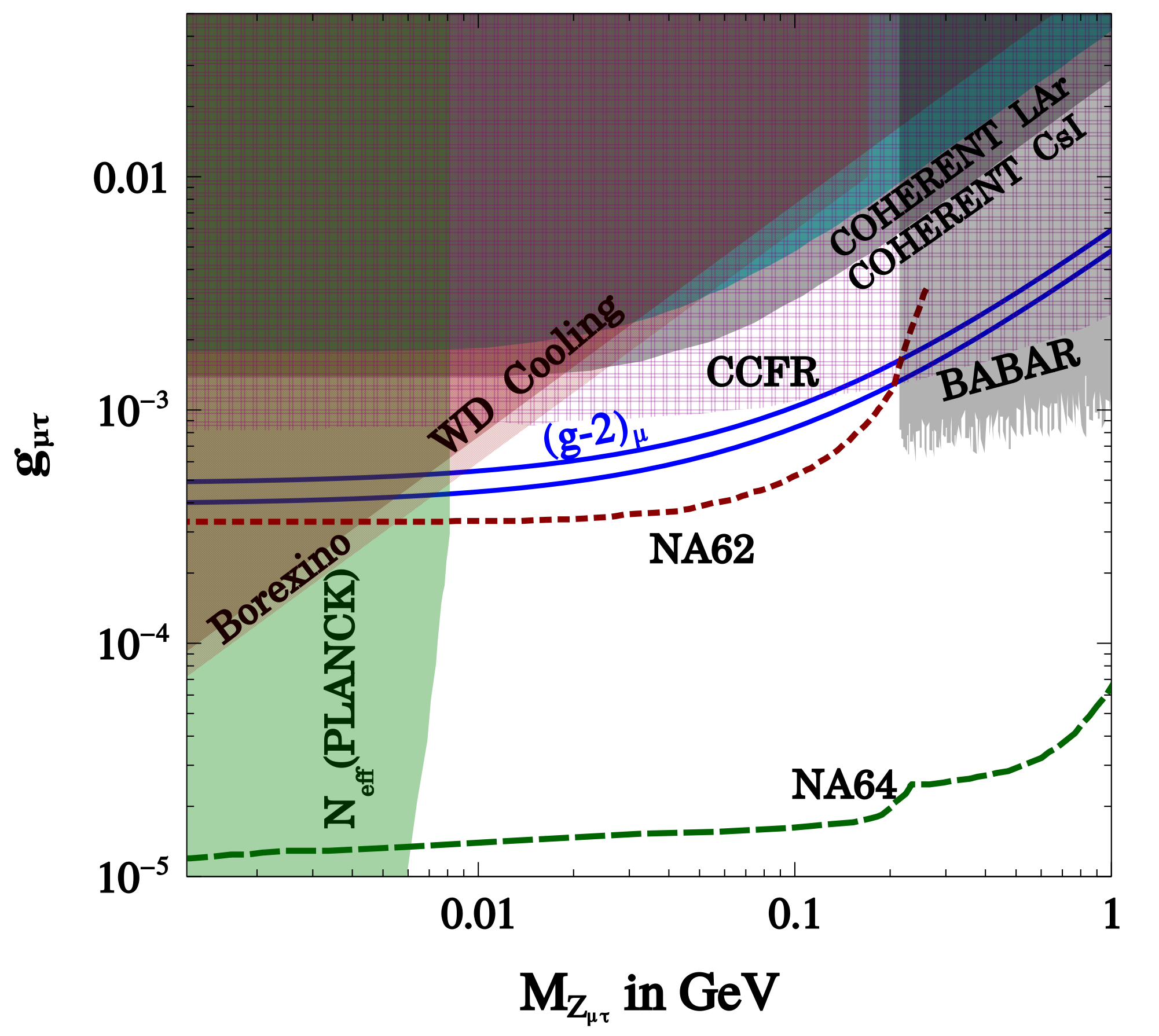}
		\caption{Parameter space satisfying correct $(g-2)_\mu$ in the plane of $g_{\mu\tau}-M_{Z_{\mu\tau}}$. See text for details of various constraints imposed on the parameter space.}
		\label{fig:muong2}
	\end{figure}
	
	In Fig.~\ref{fig:muong2}, we present the parameter space that satisfies the muon $g-2$ in the $g_{\mu \tau}$ - $M_{Z_{\mu\tau}}$ plane. This parameter space is constrained by various exclusion limits from different experiments, namely, CCFR \cite{Altmannshofer:2014pba}, COHERENT \cite{Akimov:2017ade, Akimov:2020pdx}, BABAR \cite{TheBABAR:2016rlg}. The neutrino trident constraint from CCFR is shown by a magenta-colored mesh since some
	backgrounds have not been properly taken into account~\cite{Krnjaic:2019rsv,Amaral:2021rzw}.  Exclusion zones from astrophysical bounds, specifically related to the cooling of white dwarfs (WD) \cite{Bauer:2018onh, Kamada:2018zxi}, are shown by the upper-left triangular region of Fig.~\ref{fig:muong2}. Additionally, constraints from Borexino~\cite{Bellini:2011rx}, as discussed in ~\cite{Gninenko:2020xys}, rule out a slightly larger portion compared to WD cooling constraints. Cosmological considerations of effective relativistic degrees of freedom \cite{Aghanim:2018eyx, Kamada:2018zxi, Ibe:2019gpv, Escudero:2019gzq} have effectively eliminated the possibility of very light $Z_{\mu\tau}$. This is due to the fact that delayed decay of light gauge bosons into SM leptons after the standard neutrino decoupling temperature leads to an increase in effective relativistic degrees of freedom $N_{\rm eff}$, tightly constrained by cosmic microwave background observations. Dashed lines in the figure represent the future sensitivities of NA62 \cite{Krnjaic:2019rsv} and NA64 \cite{Gninenko:2014pea, Gninenko:2018tlp} experiments. Clearly, a small region of parameter space offering a potential explanation for the muon $(g-2)$ and allowed from other experimental constraints remains verifiable in the near future. Similar observations have been made in previous studies on the minimal $L_{\mu}-L_{\tau}$ model \cite{Borah:2020jzi, Zu:2021odn, Amaral:2021rzw, Zhou:2021vnf, Borah:2021jzu, Borah:2021mri, Holst:2021lzm, Singirala:2021gok, Hapitas:2021ilr}.
	
\begin{figure}[h!]
		\centering
		\includegraphics[scale=0.25]{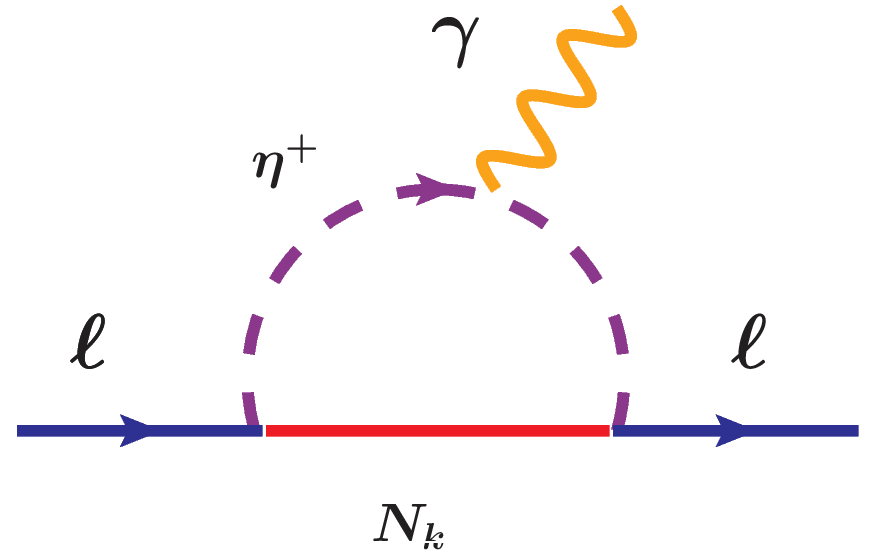}
		\caption{Contribution to $(g-2)_\mu$ from $\eta^+$ and RHN in the loop.}
		\label{fig:scotog2}
	\end{figure}

	It should also be noted that in the scotogenic version of the $L_{\mu}-L_{\tau}$ model, an extra one-loop diagram can also contribute to the anomalous magnetic moment. This diagram involves the charged component of the scalar doublet, denoted as $\eta^+$, and the right-handed neutrino $N_{k}$ in the loop, as shown in Fig.~\ref{fig:scotog2}. However, this diagram can yield negative contributions to the anomalous magnetic moments of the muon. 
	
	This contribution from the charged scalar loop is given by \cite{Queiroz:2014zfa, Jana:2020joi}
	\begin{equation}
	\Delta a_\mu =\sum_k -\frac{m^2_\mu}{8\pi^2 M^2_{\eta^+}} \lvert Y_{\mu k} \rvert^2 f(M^2_{k}/M^2_{\eta^+}),\
	\end{equation}
	where 
	\begin{equation}
	f(x)=\frac{1-6x+3x^2+2x^3-6x^2\log{x}}{12 (1-x)^4}.\
	\label{loop1}
	\end{equation}
	
	Incorporating this contribution into the overall $(g-2)_\mu$ emphasizes the necessity for the positive contribution from the vector boson loop, surpassing that from the charged scalar loop. This guarantees the overall positivity of the muon $(g-2)$ in line with experimental observations. As illustrated in Fig.~\ref{fig:muong2}, there remains a parameter space not constrained by experiments, allowing for a larger positive contribution to the muon $(g-2)$ from the vector boson loop. Therefore, even if the $\eta^+$ loop contributes negatively to the muon $(g-2)$, the observed muon $(g-2)$ can be explained by a combination of positive and negative contributions within the scotogenic $L_{\mu}-L_{\tau}$ model.

	\subsection{Lepton flavour violation}
	Charged lepton flavor-violating (CLFV) decays have been considered to be promising probes of physics beyond the SM. If we consider only the SM particle content with massive neutrinos, these processes occur only at the one-loop level and are significantly suppressed due to the tiny masses of neutrinos, keeping them well beyond the reach of current and future experimental sensitivities \cite{MEGII:2023ltw}. Consequently, any future detection of CLFV decays, such as $\mu \rightarrow e \gamma$, would serve as a distinctive indicator of physics beyond the SM. Within our model, a new one-loop contribution to CLFV arises from the charged component of $\eta$ and right-handed neutrinos in the loop, as illustrated in Fig.~\ref{fig:scotog2}. The branching ratio for the process $\mu \rightarrow e \gamma$ can be computed as \cite{Lavoura:2003xp, Toma:2013zsa}
	
	\begin{align}
	{\rm Br} (\mu \rightarrow e \gamma) =\frac{3 (4\pi)^3 \alpha}{4G^2_F} (\lvert A_{e\mu}^M \rvert^2+\lvert A_{e\mu}^E \rvert^2) {\rm Br} (\mu \rightarrow e \nu_{\mu} \overline{\nu_e}).
	\end{align}
	
	Here, $A_{e\mu}^{M, E}$ are the dipole form factors that are provided in Appendix \ref{app:lfvloop}. The most recent constraint from the MEG collaboration is $\text{Br}(\mu \rightarrow e \gamma) < 3.1 \times 10^{-13}$ at a $90\%$ CL \cite{MEGII:2023ltw}. We use this bound to constrain our parameter space to keep our analysis consistent with the desired phenomenology discussed above.  
	
	\section{Dark matter phenomenology}
	\label{sec:darkmatter}
	The model we consider here has two potential DM candidates, given that both the inert doublet scalar and the right-handed singlet fermions have odd parity under the $Z_2$ symmetry, ensuring their stability. Since the doublet scalar $\eta$ plays a pivotal role in addressing the $W$-mass anomaly and explaining the 95 GeV excess, the scalar mass spectrum, and couplings are already tightly constrained by these requirements. Consequently, our focus shifts to the study of the fermionic DM phenomenology within this framework.
	
	In our scenario, the lightest among all the odd-sector particles, denoted as $N_1$, assumes the role of dark matter. The thermal freeze-out mechanism is employed to realize the dark matter relic. Since $N_1$ is a singlet under the SM gauge symmetry, its production mechanism is intricately tied to its Yukawa couplings with scalars and fermions of the model as well as the $U(1)_{L_\mu-L_\tau}$ gauge couplings. While $U(1)_{L_\mu-L_\tau}$ gauge coupling is restricted to a tiny window from $(g-2)_\mu$ criteria, a large Yukawa coupling with SM leptons can be achieved through Casas-Ibarra parametrization by tuning $\lambda_5$ to smaller values. We have used $\lambda_5$ in the range $\lambda_5 \in [3\times10^{-8},10^{-5}]$ which gives the Yukawas in the range $Y_{\alpha k} \in [10^{-4},10^{-2}$]. In this WIMP realization of fermion singlet DM, the processes contributing to the relic density via thermal freeze-out fall into three categories: (a) DM particles annihilate to SM particles or $Z_{\mu\tau}$ through the \textit{t} or \textit{u}-channel processes; (b) DM particles annihilate into SM particles and $Z_{\mu\tau}$ through s-channel scalar mediation; (c) the coannihilations among the DM particle and the next lightest dark sector particles, as well as the annihilations of the coannihilation partners into SM sectors and $Z_{\mu\tau}$. 
	The relic density of DM has been computed using \texttt{micrOMEGAS}~\cite{Belanger:2013oya}, which takes into account all relevant annihilation and coannihilation processes.
	
	\begin{figure}[h!]
		\centering
		\includegraphics[scale=0.45]{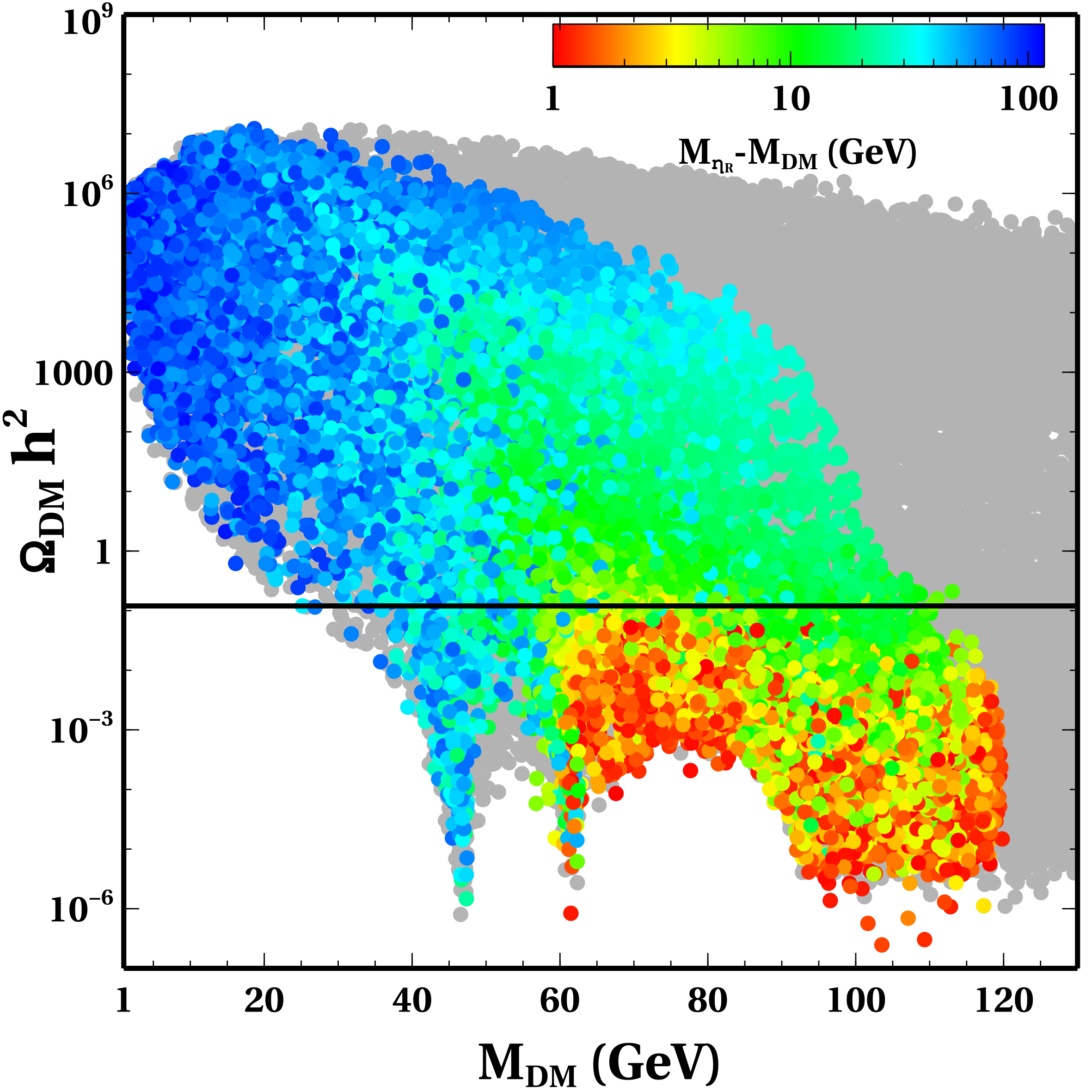}
		\caption{DM relic density as a function of DM mass.}
		\label{fig:relicvsdmmass}
	\end{figure}
	
	We have shown the variation of DM relic density as a function of DM mass in Fig. \ref{fig:relicvsdmmass}.  For the scan, the Yukawa couplings are calculated using the Casas-Ibarra parametrization, given in Eq. \eqref{eq:Yuk}. The right handed neutrinos are varied in the range : $M_{N_1}\equiv M_{\rm DM}\in[1, 150]$ GeV, $M_{N_2}=M_{N_3}$=$M_{N_1}+100$ GeV while all other parameters are varied as discussed in section~\ref{sec:95gev}.
	
	The grey region in Fig. \ref{fig:relicvsdmmass} is excluded by imposing the constraints from W-mass anomaly, 95 GeV signal strength, and SM Higgs signal strength, as discussed earlier.
	The color coding shows the mass splitting between the DM and the next to lightest stable particle (NLSP), $\eta_{R}$. This mass splitting plays an important role in determining the processes involved in relic generation. When the mass splitting is large, the coannihilations are negligible, and the scalar-mediated annihilation channels dominate. The sharp dips around $47.5$ GeV and $62.5$ GeV are due to the scalar resonances corresponding to 95 GeV scalar $H_2$ and SM Higgs $H_1$.  When that mass splitting between DM and NLSP decreases, the contributions of co-annihilation processes also become significant in addition to the annihilation processes, and consequently, the relic decreases, as seen from Fig. \ref{fig:relicvsdmmass}. Clearly, we can see that correct relic density\cite{Planck:2018vyg} can be achieved for DM mass in the range $M_{\rm DM} \in [25,110]$ GeV.
	
\begin{figure}[h!]
		\centering
		\includegraphics[scale=1.1]{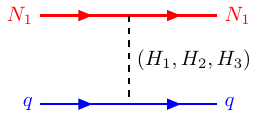}
		\includegraphics[scale=1.1]{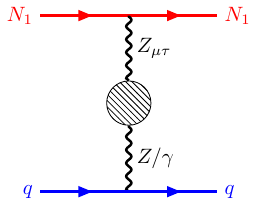}
		\caption{DM-nucleon scattering mediated by the scalars (top) and vector bosons (bottom).}
		\label{fig:DDdiag}
	\end{figure}

	Since DM is of WIMP type, we have the possibility of probing it at direct detection experiments. Here, the DM-nucleon scattering is possible via scalar and vector boson-mediated interactions, as shown in Fig. \ref{fig:DDdiag}.
	
	The spin-independent (SI) DM-nucleon cross-section via scalar mediation is given as \cite{Ellis:2008hf}
	\begin{equation}
	\sigma^{\rm SI}_{\rm scalar}=\frac{\mu_r^2}{\pi A^2}[\mathcal{M}]^2,\
	\end{equation}
	where $\mu_r=\frac{M_{\rm DM} m_n}{M_{\rm DM}+m_n}$ is the reduced mass, $m_n$ is the nucleon (proton or
	neutron) mass. $\mathcal{M}$ is the amplitude corresponding to the scalar mediated diagram shown in Fig. \ref{fig:DDdiag}. This amplitude is given as,
	\begin{eqnarray}
	\mathcal{M}=  \mathcal{Z} f_{p} + (A-\mathcal{Z}) f_{n},\
	\end{eqnarray}
	where $A$ is the mass number of target nucleus, $\mathcal{Z}$ is the atomic number of target nucleus. The $f_{p}$, and $f_{n}$ are the interaction strengths of proton and neutron with DM, respectively, and are given as,
	\begin{equation}
	f^i_{p,n}=\sum_{q=u,d,s} f_{T_q}^{p,n} \alpha^i_{q} \frac{m_{p,n}}{m_q}+\frac{2}{27} f_{T_G}^{p,n} \sum_{q=c,t,b}\alpha^i_{q} \frac{m_{p,n}}{m_q},\
	\end{equation}
	where,
	\begin{equation}
	\alpha^i_{q}=Y_{H_i}\times\frac{m_q}{v}\times\bigg(\frac{1}{m^2_{H_i}}\bigg).\
	\end{equation}
	Here the values of $f_{T_q}^{p,n}$, $f_{T_G}^{p,n}$ can be found in \cite{Ellis:2000ds}.
	In the above equation, $Y_{H_i}$ are the effective couplings between DM and the scalars ($H_1$, $H_2$, and $H_3$) and are given as,
	\begin{eqnarray}
	Y_{H_1}&=\sin{\beta_{12}}\cos{\beta_{12}}(-c_{12} s_{12} s_{23}-s_{23}c_{23})Y_{e\mu}+\nonumber\\& \frac{1}{\sqrt{2}}\sin^2{\beta_{12}}(s_{12}s_{23}-c_{12}s_{13}c_{23})Y_\mu,\
	\end{eqnarray}
	\begin{eqnarray}
	Y_{H_2}&=\sin{\beta_{12}}\cos{\beta_{12}}(-s_{12}s_{13}s_{23}+c_{12}c_{23})Y_{e\mu}+\nonumber\\& \frac{1}{\sqrt{2}}\sin^2{\beta_{12}}(-c_{12}s_{23}-s_{12}s_{13}c_{23})Y_\mu,\
	\end{eqnarray}
	\begin{eqnarray}
	Y_{H_3}&=\sin{\beta_{12}}\cos{\beta_{12}}(c_{13}s_{23})Y_{e\mu}+\nonumber\\&\frac{1}{\sqrt{2}}\sin^2{\beta_{12}}(c_{13}c_{23})Y_\mu,\
	\end{eqnarray}
	where $s_{ij}$ and $c_{ij}$ are the scalar mixing angles and $\beta_{12}$ is the RHN mixing angle as discussed in Section \ref{sec:95gev} and Appendix \ref{app:rhnmixing} respectively. The expressions for $Y_{e\mu}$ and $Y_\mu$ can be found in Appendix \ref{app:rhnmixing}.
	
	\begin{figure}[h!]
		\centering
		\includegraphics[scale=0.45]{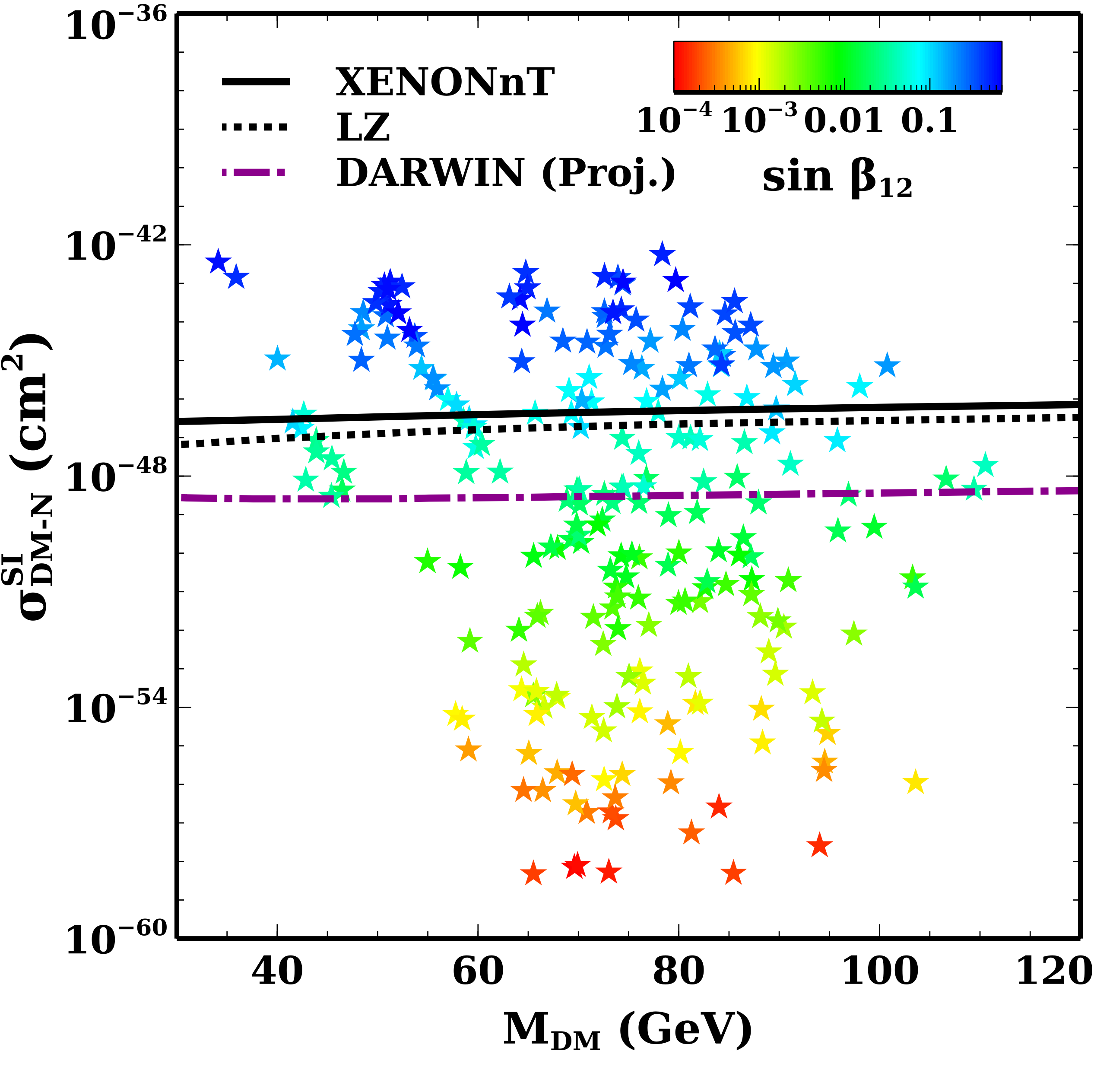}
		\caption{Spin-independent direct detection cross-section as a function of DM mass for the points satisfying correct relic density.}
		\label{fig:dd}
	\end{figure}
	In addition to the scalar mediation process responsible for DM-nucleon scattering, a vector portal interaction can also occur due to the kinetic mixing between the $Z_{\mu\tau}$ with $Z$ boson or photon in the SM. With the proper choice of the basis, it is possible to rewrite the Lagrangian such that the gauge fields' kinetic terms are
		in the canonical form and the kinetic mixing term is removed. The way to achieve this is not unique as shown in~\cite{Pan:2018dmu}, and it is possible to redefine the gauge fields such that photon has no interaction with $Z_{\mu\tau}$ and hence with the dark gauge current.  However, it is important to note that in a gauged $L_\mu-L_\tau$ setup that we considered here,  even though the kinetic mixing at tree level can be assumed to be zero or the kinetic mixing between $Z_{\mu\tau}$ and photon can be removed by redefinition of the gauge fields, it is always induced at the one-loop level with the particles charged under
		both $U(1)_Y$ and $U(1)_{L_\mu-L_\tau}$ running in the loop. Thus, the kinetic mixing parameter can not be tuned arbitrarily and its one-loop value  is given by \cite{Borah:2021jzu,Araki:2017wyg}
		\begin{equation}
		| \epsilon| \simeq \frac{g g_{\mu\tau}}{16 \pi^2}\log \left(\frac{m_\mu}{m_\tau}\right) \simeq \frac{g_{\mu\tau}}{70} .\
		\end{equation}
	
	The corresponding Feynman diagram for DM-nucleon scattering is depicted in the lower panel of Fig.~\ref{fig:DDdiag}. 
	Due to the nature of gauge coupling of DM, the vector-mediated scattering cross-section is either velocity-suppressed or spin-dependent, both of which lead to weak constraints. The spin-independent cross-section for the vector boson($Z-Z_{\mu\tau}$) mediated process is given by~\cite{Chao:2019lhb}:
	
	\begin{eqnarray}
		\sigma^{\rm SI}_{\rm vector}= \frac{ 2\mu^2_r}{\pi M^4_{Z_{\mu\tau}}} \Theta^2~ g^2 g^2_{\mu\tau} \epsilon^2 v_{_{\rm DM}}^2,\ \label{eq:SI_vector}
		\end{eqnarray}
	Here, $v_{_{\rm DM}}$ represents the velocity of the dark matter particle  at present epoch and $\Theta$ is the effective coupling of the dark sector interaction $\Bar{N_1}\gamma^\mu (Z_{\mu\tau})_{\mu}N_{1}$ given in Eq \ref{eq:n1n1zmt}. A similar expression can be obtained for the ($\gamma-Z_{\mu\tau}$) exchange process by replacing g with $\sqrt{4\pi\alpha_e}$ in Eq (\ref{eq:SI_vector}), where $\alpha_e$ is the fine structure constant. 
	
	Furthermore, spin-dependent DM-nucleon scattering can occur via vector boson mediation, with the cross-section given by
	
	\begin{eqnarray}
		\sigma^{\rm SD}= \frac{4 \mu^2_r}{\pi M^4_{Z_{\mu\tau}}} \Theta^2~ g^2 g^2_{\mu\tau} \epsilon^2\bigg(\sum_{q}\Delta^N_q\bigg)^2J_N(J_N+1).\nonumber\\ \label{eq:SDDD}
		\end{eqnarray}

	Here, $J_N$ represents the angular momentum of the nucleon, and $\Delta^N_q$ denotes the spin fraction of quarks, as detailed in~\cite{Agrawal:2010fh}. And our parameter space remains completely safe from the existing constraints~\cite{LZ:2022lsv}. So, for our analysis, we exclusively consider the spin-independent DM-nucleon scattering, which is tightly constrained by terrestrial DM search experiments. We also note that the $\sigma^{\rm SI}_{\rm vector}$ always remains sub-dominant as compared to $\sigma^{\rm SI}_{\rm scalar}$ because of the velocity suppression. For all practical purposes, we set $\sigma^{\rm SI}_{\rm vector}=0$ and $\sigma^{\rm SD}=0$ in Eq(\ref{eq:SI_vector}) by considering a typical choice of the texture of the RHN mass matrix as given in Eq. (\ref{eq:RHNmass}), which predicts $\Theta=0$. We note that this typical choice simplifies our calculations without a drastic change of the phenomenological consequences achieved in other sections.
	
	We have shown the SI direct detection cross-section as a function of DM mass in Fig. \ref{fig:dd} where the color coding depicts the RHN mixing angle ($\beta_{12})$. Clearly, the existing experimental bounds from the XENON1T \cite{Aprile:2018dbl} and LZ \cite{LZ:2022lsv} put an upper bound on the mixing angle $\sin{\beta_{12}}< 0.04$. However, future experiments like DARWIN \cite{DARWIN:2016hyl} has the potential to probe this mixing angle down to $\sin{\beta_{12}}= 0.02$. Here, it is worth mentioning that, in addition to the tree-level DM-nucleon scattering process shown in Fig.~\ref{fig:DDdiag}, it can also arise at one-loop level with the doublet scalars in the loop as studied in~\cite{Ibarra:2016dlb}. However, because of loop suppression, it always remains sub-dominant as compared to the tree-level cross-section. 
	
	\section{Conclusion}
	\label{sec:conclusion}
	We have studied the possibility of finding a common origin of three particle physics anomalies, namely, the muon $(g-2)$ anomaly, CDF-II W-mass anomaly, and CMS 95 GeV excess within the framework of scotogenic $L_\mu-L_\tau$ model which also explains the origin of light neutrino mass and dark matter in the Universe. While the lightest among the singlet right-handed neutrinos responsible for generating light neutrino mass plays the role of DM, the $L_\mu-L_\tau$ gauge boson explains the muon $(g-2)$ anomaly. The inert scalar doublet can lead to the required enhancement of W-boson mass via radiative corrections. While a neutral scalar formed out of the singlet scalars responsible for $L_\mu-L_\tau$ gauge symmetry breaking with a tiny admixture of the SM Higgs can play the role of the 95 GeV scalar, the required branching ratio for diphoton decay mode of this new light scalar together with that of the SM Higgs forces one to include one more inert Higgs doublet such that they can give rise to additional one-loop contributions to these decay widths without changing rest of the phenomenology. While this new scalar doublet does not play any additional role in our setup, it can be motivated from neutrino mass point of view if we have only one right-handed neutrino. The parameter space consistent with all the requirements remains verifiable at ongoing and near-future experiments like dark matter direct detection, dark photon searches, charged lepton flavor violation, as well as colliders.

	\noindent
	\acknowledgments
	The work of DB is supported by the Science and Engineering Research Board (SERB), Government of India grant MTR/2022/000575.  SM acknowledges the financial
	support from the National Research Foundation of Korea grant 2022R1A2C1005050. PKP would like to acknowledge the Ministry of Education, Government of India, for providing financial support for his research via the Prime Minister’s Research Fellowship (PMRF) scheme. The work of NS is supported by the Department of Atomic Energy-Board of Research in Nuclear Sciences, Government of India (Ref. Number: 58/14/15/2021- BRNS/37220).
	\appendix
	\section{RHN Masses and Mixing}\label{app:rhnmixing}
	From Eq \ref{mass_mat} the texture of the right-handed neutrino mass matrix in the flavor basis i.e. $(N_e,N_\mu, N_\tau)^T$ is given by,
		\begin{eqnarray}
		M_R=\left(
		\begin{array}{ccc}
		a & b  & c \\
		b & d &e \\
		c &e &f \\
		\end{array}
		\right)\label{eq:RHNmass1}
		\end{eqnarray}
		Assuming $M_R$ to be real, it can be diagonalized by using an orthogonal matrix, $O=O_{12}(\beta_{12}).O_{13}(\beta_{13}).O_{23}(\beta_{23})$
		\begin{equation}
		O=\left(
		\begin{array}{ccc}
		\text{c12} \text{c13} & \text{c23} \text{s12}-\text{c12} \text{s13} \text{s23} & \text{c12} \text{c23}
		\text{s13}+\text{s12} \text{s23} \\
		-\text{c13} \text{s12} & \text{c12} \text{c23}+\text{s12} \text{s13} \text{s23} & \text{c12}
		\text{s23}-\text{c23} \text{s12} \text{s13} \\
		-\text{s13} & -\text{c13} \text{s23} & \text{c13} \text{c23} \\
		\end{array}
		\right)\\
		\label{eq:Dmtrasf}
		\end{equation}
		to get the mass-eigen state $N_{1},N_{2},N_{3}$ such that the lightest one, $N_1$ is the DM. Here $\cos\beta_{ij}=cij$, and $\sin\beta_{ij}=sij$. Using Eq (\ref{eq:Dmtrasf}) and Eq (\ref{eq:yuklag}), the effective coupling of $\bar{N_1}\gamma^{\mu}(Z_{\mu\tau})_{\mu}N_1$ can be written as \begin{eqnarray}
		\Theta&=&(\cos\beta_{23}\sin\beta_{12}-\cos\beta_{12}\sin\beta_{13}\sin\beta_{23})^2\nonumber\\&-&(\sin\beta_{23}\sin\beta_{12}-\cos\beta_{12}\sin\beta_{13}\cos\beta_{23})^2\label{eq:n1n1zmt}  
		\end{eqnarray}
		For simplicity, we choose the texture of the RHN mass matrix to be 
	\begin{eqnarray}
		M_R=\left(
		\begin{array}{ccc}
		a & b  & b \\
		b & c &d \\
		b &d &c \\
		\end{array}
		\right)\label{eq:RHNmass}
		\end{eqnarray}
	
	where $a=M_{ee},~ b=(Y_{e\mu}=Y_{e\tau})v_1,~ c=(Y_{\mu}=Y_\tau)v_2,~ d=M_{\mu\tau}$. Assuming $a,b,c,d$ to be real in Eq(\ref{eq:RHNmass}), $M_R$ can be diagonalized by using an orthogonal matrix, $U=U_{12}(\beta_{12}).U_{13}(\beta_{13}=0).U_{23}(\beta_{23}=\frac{\pi}{4})$
		\begin{eqnarray}
		U=\left(
		\begin{array}{ccc}
		\cos\beta_{12} & \frac{\sin\beta_{12}}{\sqrt{2}}  & \frac{\sin\beta_{12}}{\sqrt{2}} \\
		-\sin\beta_{12} & \frac{\cos\beta_{12}}{\sqrt{2}} &\frac{\cos\beta_{12}}{\sqrt{2}} \\
		0 &-\frac{1}{\sqrt{2}} &\frac{1}{\sqrt{2}} \\
		\end{array}
		\right),
		\end{eqnarray}
		to get the mass eigen-states $N_{1}, N_{2}, N_{3}$ with $N_1$ being the DM.
		Using Eq(\ref{eq:Dmtrasf}) and Eq(\ref{eq:yuklag}), the effective coupling of $\bar{N_1}\gamma^{\mu}(Z_{\mu\tau})_{\mu}N_1$ is found to be $\Theta=0$.
	
	The couplings and the mass parameters can be expressed in terms of the physical RHN masses, $M_{N_1}, M_{N_2}, M_{N_3}$ and mixing angle, $\beta_{12}$ as,
	\begin{eqnarray}
	M_{ee}=\cos^2\beta_{12}M_{N_1}+\sin^2\beta_{12}M_{N_2}
	\end{eqnarray}
	\begin{eqnarray}
	Y_\mu=Y_\tau&=\frac{1}{v_2}(\sin^2\beta_{12}\frac{M_{N_1}}{2}+\cos^2\beta_{12}\frac{M_{N_2}}{2}+\frac{M_{N_3}}{2})\nonumber\\
	\end{eqnarray}
	\begin{eqnarray}
	M_{\mu\tau}&=\frac{1}{v_2}(\sin^2\beta_{12}\frac{M_{N_1}}{2}+\cos^2\beta_{12}\frac{M_{N_2}}{2}-\frac{M_{N_3}}{2})
	\end{eqnarray}
	\begin{eqnarray}
	Y_{e\mu}=Y_{e\tau}=\frac{1}{v_1}\big(\frac{M_{N_1}}{\sqrt{2}}-\frac{M_{N_2}}{\sqrt{2}}\big)\cos\beta_{12} \sin\beta_{12}
	\end{eqnarray}
	\section{Physical masses of inert doublet scalar components}
	\label{app:idmmass}
	
	The $Z_2$-odd scalars can be written in component form as,
	\begin{eqnarray}
	\eta=\left(
	\begin{array}{c}
	\eta^+ \\
	\frac{\eta_{R}+i\eta_{I}}{\sqrt{2}}\\
	\end{array}
	\right)
	\end{eqnarray}
	The mass squared of the charged and neutral components are given as,
	\begin{eqnarray}
	M_{\eta_{R}}^2&=m_\eta^2+\frac{v^2}{2}(\lambda_3+\lambda_4+\lambda_5)+\frac{v_1^2}{2}\lambda_{\eta \Phi_1} +\frac{v_2^2}{2}\lambda_{\eta \Phi_2}\label{eq:metar}\nonumber\\
	\end{eqnarray}
	\begin{eqnarray}
	M_{\eta_{I}}^2&=m_\eta^2+\frac{v^2}{2}(\lambda_3+\lambda_4-\lambda_5)+\frac{v_1^2}{2}\lambda_{\eta \Phi_1} +\frac{v_2^2}{2}\lambda_{\eta \Phi_2}\label{eq:metai}\nonumber\\
	\end{eqnarray}
	\begin{eqnarray}
	M_{\eta^+}^2&=m_\eta^2+\frac{v^2}{2}\lambda_3+\frac{v_1^2}{2}\lambda_{\eta \Phi_1}+\frac{v_2^2}{2}\lambda_{\eta \Phi_2}\label{eq:metap}
	\end{eqnarray}
	\section{Loop functions}
	\label{app:loop}
	The loop functions involved in the calculation of $\Gamma(h_i\to
	\gamma\gamma)$ are given
	by~\cite{Djouadi:2005gi,Staub:2016dxq}
	\begin{align}
	A_0 &=-[\tau -f(\tau)]/\tau^2 \, , \nonumber \\
	A_{1/2} &= 2[\tau + (\tau-1)f(\tau)]/\tau^2 \, , \\
	A_{1} &=-[2\tau^2+3\tau + 3(2\tau -1)f(\tau)]/\tau^2 \, , \nonumber
	\end{align}
	where the function $f(\tau)$ is defined as
	\begin{eqnarray}
	f(\tau)=\left\{ \begin{array}{lc}
	\arcsin^2 \sqrt{\tau}; &   \tau \leq 1 \\
	\\ -\frac{1}{4}\left[\log \frac{1+\sqrt{1-\tau^{-1}}}{1-\sqrt{1-\tau^{-1}}}-i\pi\right]^2; &   \tau > 1 
	\end{array}
	\right.\nonumber\\
	\end{eqnarray}
	\section{CLFV}
	\label{app:lfvloop}
	\begin{eqnarray}
	I^{(\pm)_1(\pm)_2}_{k}=\int d^3X\frac{x(y+(\pm)_1z \frac{m_{e}}{m_{\mu}}+(\pm)_2\frac{M_{k}}{m_{\mu}})}{-xy m_{\mu}^2-xzm_e^2+(1-x)M_{\eta^+}^2+x M_k^2}\nonumber\\
	\end{eqnarray}
	
	\begin{eqnarray}
	A_{e\mu}^M=\frac{-1}{(4\pi)^2}\sum_{k}(Y_{e k}^* Y_{\mu k} I^{++}_k+Y_{e k}^* Y_{\mu k} I^{+-}_k)
	\end{eqnarray}
	\begin{eqnarray}
	A_{e\mu}^E=\frac{-i}{(4\pi)^2}\sum_{k}(-Y_{e k}^* Y_{\mu k} I^{-+}_k-Y_{e k}^* Y_{\mu k} I^{--}_k)
	\end{eqnarray}
	
	\vfill
	\eject

\end{document}